\documentclass[fleqn,10pt]{wlscirep}
\usepackage[utf8]{inputenc}
\usepackage[T1]{fontenc}
\usepackage{booktabs}
\title{Commercial CMOS Process for Quantum Computing: Quantum Dots and Charge Sensing in a 22~nm Fully Depleted Silicon-on-Insulator Process}

\author[1]{S.V.~Amitonov, A.~Aprà, M.~Asker, B.~Barry, I.~Bashir, P.~Bisiaux, E.~Blokhina, P.~Giounanlis,  P.~Hanos-Puskai, M.~Harkin, I.~Kriekouki, D.~Leipold, M.~Moras,   C.~Power, N.~Samkharadze, A.~Sokolov, D.~Redmond, C.~Rohrbacher, X.~Wu}

\affil[1]{Equal1 Laboratories}

\begin{abstract}
  Confining electrons or holes in quantum dots formed in the channel of industry-standard fully depleted silicon-on-insulator CMOS structures is a promising approach to scalable qubit architectures. In this communication, we present measurement results of a commercial nanostructure fabricated using the GlobalFoundries 22FDX\textsuperscript{TM} industrial process. We demonstrate here that quantum dots are formed in the device channel by applying a combination of a back- and gate voltages. We report our results on an effective detuning of the energy levels in the quantum dots by varying the barrier gate voltages in combination with the back-gate voltage. Given the need and importance of scaling to larger numbers of qubits, we demonstrate here the feasibility of single-electron box sensors at the edge of the quantum dot array for effective charge sensing in different operation modes -- sensing charge transitions in a single- and double quantum dots forming the quantum dot array. We also report measurement results demonstrating bias triangle pair formation and precise control over coupled quantum dots with variations in the inter-dot barrier.  The reported measurement results demonstrate the ability to control the formation and coupling of multiple quantum dots in a quantum dot array and to sense their charge state via a Single Electron Box sensor in a commercial process for the first time.  

\end{abstract}

\begin{document}

\flushbottom
\maketitle
\thispagestyle{empty}

\section*{Introduction}\label{sec:introduction}

Qubits in electrostatically defined quantum dots have emerged as a leading platform for scalable quantum computing, offering excellent coherence, small footprint and compatibility with industrial fabrication processes~\cite{elsayed_low_2024}. These quantum dots can encode quantum information using various degrees of freedom, such as the charge and spin states of individual charge carriers, as well as multi-particle and multi-dot states acting as qubits, including singlet/triplet, exchange only, flip-flop, and hybrid states~\cite{burkard_semiconductor_2023}. 

The typical structure and dimensions of electrostatically defined quantum dots closely resemble those of transistors, featuring an undoped channel to host charge carriers, a doped source for charge injection, and nanoscale gates—typically tens of nanometers in size—for controlling charge occupation within the channel. However, unlike transistors, qubit arrays require direct interaction between quantum dots formed within the same channel. This necessitates precise control over the tunnel coupling between the dots and a mechanism to read out the charge states of the multi-dot structures.

In this paper, we present the experimental characterisation of a quantum dot array (QDA) manufactured in a commercial 22~nm Fully Depleted Silicon-On-Insulator (FDSOI) process from GlobalFoundries~\cite{ong_22nm_2019, bonen_investigation_2024}. The process would allow monolithic integration of quantum dots, charge sensors and control electronics~\cite{bashir2024,staszewski2022cryogenic}. We demonstrate full electrical control over the location and coupling of quantum dots in the 22FDX\textsuperscript{TM} industrial process. An effective plunger gate is realised through variations in the backgate voltage and the barrier gate voltages depending on the region of operation~\cite{sokolov_common-mode_nodate}.
Firstly, we demonstrate the confinement, charge stability, and bias triangles (in forward and reverse bias) forming a double quantum dot structure by measuring it in transport. We then show that the barrier gate separating two quantum dots has precise control over the inter-dot coupling~\cite{lai_pauli_2011}. Furthermore, following from our previous work~\cite{bashir2020single,petropoulos2024}, we show that we can configure the same QDA into a single-lead Single Electron Box (SEB)~\cite{house2016} with one or two adjacent quantum dots, allowing us to  measure charge stability of the one (or two) quantum dots in SEB sensing mode. The latter provides a full functionality to control quantum dots and sense charge in one quantum dot array line.

\section*{System Overview and Measurement Setup\label{sec:measurement_setup}}

\begin{figure}[ht]
\centering
\includegraphics[width=0.9\linewidth]{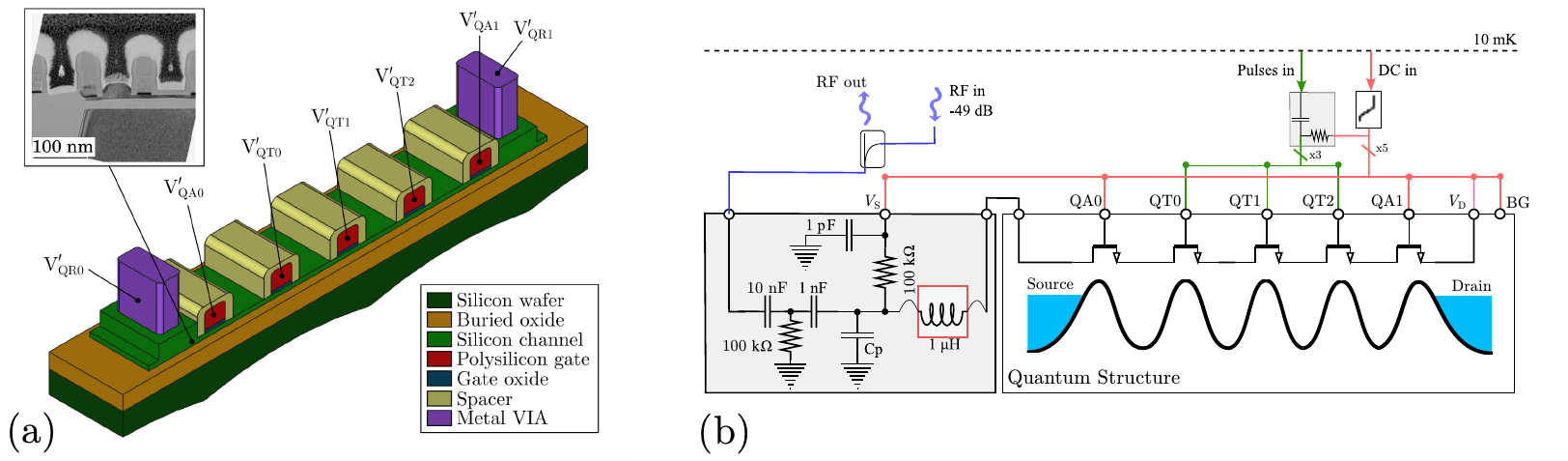}
\caption{\label{fig:exp_circuit} (a) 3D view of the five-gate quantum dot array with raised source and drain. The scanning electron microscope (SEM) image of the device shows dummy polysilicon gates that are not included in the 3D view. A backgate terminal is also available in this process but is not visible in this diagram. It connects through a metal VIA to the silicon wafer below the buried oxide. (b) A simplified experimental setup of the measurement configuration that was used in this work. The RF tank circuit is connected to the source electrode on the device. The superconducting NbTiN inductor is glued and wirebonded on the PCB. All barrier gates are connected to on-PCB bias-tees for pulsing. The bias tees and tank circuit are located on the same PCB but have been separated here for clarity. The base temperature of the dilution refrigerator is $10~\text{mK}$.}
\end{figure}

The device investigated in this study is fabricated using GlobalFoundries' 22~nm Fully Depleted Silicon-On-Insulator (FD-SOI) technology. As illustrated in Fig.\ref{fig:exp_circuit}(a), the device consists of a raised source and drain, and five electrostatic gates and it serves as an initial test structure for future highly scalable QDAs based on industry-standard FD-SOI processes. 
As can be seen in this figure, unlike other QDAs with quantum dots controlled by plunger gates\cite{burkard_semiconductor_2023}, this device admits indirect electrical control over quantum dots forming between the polysilicon barrier gates. The fabrication of the device does not require modification of the standard photolithography process\,\cite{wu_evolution_2020, hou_optical_2021, park_patterning_2023}. We emphasise that the most pertinent feature of the standard processing that we must adhere to is standard sizing and pitch, which places strict constraints on the design of quantum dot devices in photolithography-fabricated devices. The gates QA0 and QA1 act as spacers between the heavily doped source and drain and the quantum dots. The terminal voltages and potentials used in the control of this device are shown in Tab.~\ref{tab:terminalvoltages}. Full modelling and simulation of the QDA discussed here is provided in \cite{sokolov_common-mode_nodate}, along with further experimental characterisation.

\begin{table}[h!]
\centering
\begin{tabular}{@{}ll|llc@{}}
\toprule
Voltage & Description &  Voltage & Description & Equation \\
\midrule
$V'_\text{QR0}$ & Source potential & $V_\text{CM}$  & Common-mode voltage        & $\left( V'_\text{QR0} + V'_\text{QR1} \right)/2$ \\
$V'_\text{QR1}$ & Drain potential  & $V_\text{DS}$  & Drain-to-source voltage    & $V'_\text{QR1} - V'_\text{QR0}$ \\
$V'_\text{QA0}$ & QA0 potential    & $V_\text{QA0}$ & QA0 potential w.r.t source & $V'_\text{QA0} - V'_\text{QR0}$ \\
$V'_\text{QA1}$ & QA1 potential    & $V_\text{QA1}$ & QA1 potential w.r.t source & $V'_\text{QA1} - V'_\text{QR0}$ \\
$V'_\text{QT0}$ & QT0 potential    & $V_\text{QT0}$ & QT0 potential w.r.t source & $V'_\text{QT0} - V'_\text{QR0}$ \\
$V'_\text{QT1}$ & QT1 potential    & $V_\text{QT1}$ & QT1 potential w.r.t source & $V'_\text{QT1} - V'_\text{QR0}$ \\
$V'_\text{QT2}$ & QT2 potential    & $V_\text{QT2}$ & QT2 potential w.r.t source & $V'_\text{QT2} - V'_\text{QR0}$ \\
\bottomrule
\end{tabular}
\caption{Terminal voltages and potentials used in the control of the five-gate quantum dot array in this paper.}\label{tab:terminalvoltages}%
\end{table}

The experiments were carried out in a temperature controlled dilution refrigerator with a temperature ranging from 10 mK to 700 mK. The device chip is interfaced by flip-chip mounting on a cryo-compatible laminate which is then connected onto a PCB. DC voltages are generated using a QDevil QDAC-II\cite{noauthor_quantum_nodate} high-precision low-noise digital-to-analog converter and filtered at base temperature.  The barrier gates QT0, QT1, and QT2 are connected to on-PCB bias tees to enable fast pulsing. A resonator is connected to the source electrode of the device which, with the parasitic capacitance of the PCB, allows rf-reflectometry detection\cite{Vigneau2023} at a resonance frequency of 75.5 MHz. The resonator consists of a superconducting meander inductor in NbTiN (L = 1~$\mathrm{\mu H}$) glued to the PCB and connected through wire bonding. The rf-tone used for readout is sent through an attenuated coax line and generated using an OPX+. The RF carrier is reflected by the tank circuit on the PCB sample board amplified by a cryogenic amplifier at 4 K and amplified again at room temperature. The signal is then digitized at 1 GS/s using the OPX+. The simplified experimental setup employed in this study is shown in Fig.~\ref{fig:exp_circuit}(b).

\section*{Measurement Results: Quantum Dot Array\label{sec:measurement_results_qda}}

\begin{figure}[ht]
    \centering
    \includegraphics[width=0.8\linewidth]{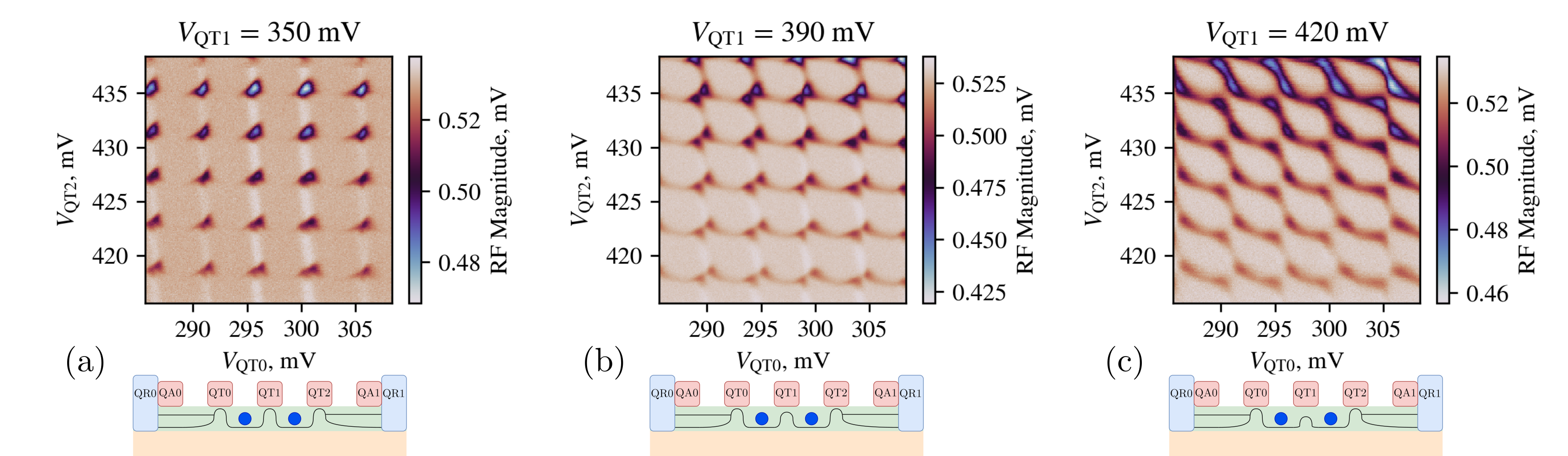}
    \caption{Variation in interdot capacitive coupling and triangle formation with central tunnel barrier. These measurements were taken at 700~mK in transport mode. Quantum dot locations are denoted with blue circles in the side views of the device. (a) Charge stability response for $V_{\text{QT1}} = 350~\text{mV}$. Bias triangle pairs are overlapping and forming at the triple points, showing clear double quantum dot formation. (b) Charge stability response for $V_{\text{QT1}} = 390~\text{mV}$. Bias triangle pairs are beginning to separate with the increased coupling between the two quantum dots. (c) Charge stability response for $V_{\text{QT1}} = 420~\text{mV}$. The response is akin to that of a single quantum dot as both dots begin to merge.}
    \label{fig:Triangles_Fig_14_19_20}
\end{figure}

We now demonstrate precise control over the inter-dot tunnel barrier at 700~mK in Fig.~\ref{fig:Triangles_Fig_14_19_20}. Two quantum dots are formed between the barrier gate pairs, QT0/QT1 and QT1/QT2. Therefore, by varying $V_{\text{QT1}}$, the voltage on the gate that separates the two quantum dots, we demonstrate control over the tunnel barrier between these two dots. These measurements are done in transport mode with $V_{\text{DS}} = 0.5~\text{mV}$, such that electron reservoirs are formed to the left and right of QT0 and QT2, respectively. 

The barrier in Fig.\ref{fig:Triangles_Fig_14_19_20}(a) is such that the two quantum dots are effectively decoupled, showing square charge stability regions~\cite{van_der_wiel_electron_2002} for $V_{\text{QT1}} = 350~\text{mV}$. Note that the cross-coupling between the barrier gates QT0 and QT2 is extremely low with the charge stability showing little to no slope in the charge transition lines. Increasing $V_{\text{QT1}} = 390~\text{mV}$ we see the triple points in the charge stability begin to separate, showing very clear bias triangle pairs in Fig.~\ref{fig:Triangles_Fig_14_19_20}(b). Finally, we then consider $V_{\text{QT1}} = 420~\text{mV}$, which begins to demonstrate the transition from a double quantum dot to a large single quantum dot between QT0 and QT2. The formation of diagonal transition lines in Fig.~\ref{fig:Triangles_Fig_14_19_20}(c) is indicative of a single quantum dot controlled by $V_{\text{QT0}}$ and $V_{\text{QT2}}$. We note the variation in the magnitude response in the charge stability measurements becomes larger for increasing $V_{\text{QT0}}$ and $V_{\text{QT2}}$. This is likely due to increased coupling from the quantum dots to the electron reservoirs, thus increasing the tunnelling rates from reservoir to dot.

\begin{figure}[h]
    \centering
    \includegraphics[width=0.45\linewidth]{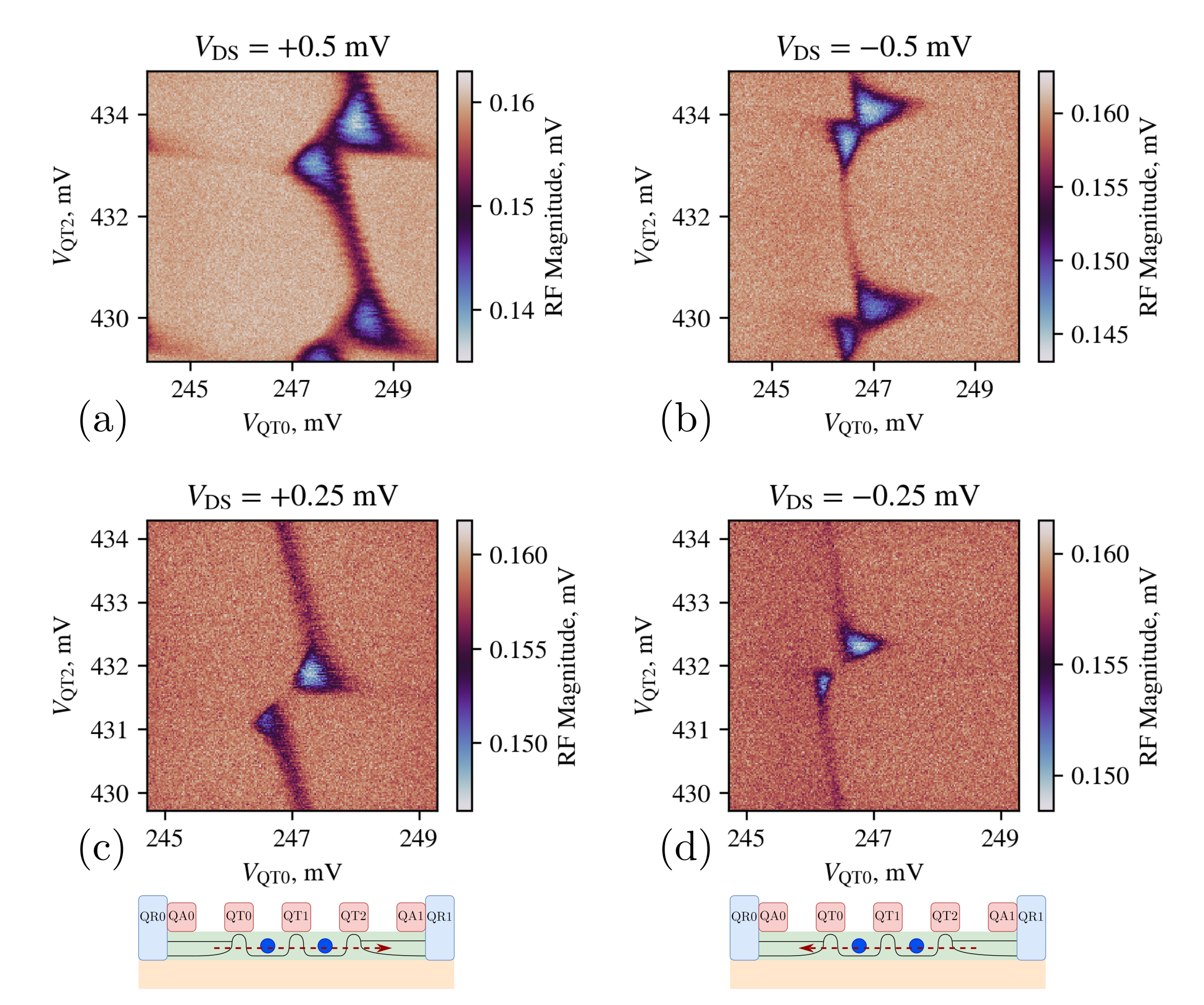}
    \caption{Measurements of specific bias triangle pairs at $70~\text{mK}$. A change in the direction of the triangles is clear with positive and negative $V_{\text{DS}}$. The size of the triangles is also proportional to the magnitude of the applied $V_{\text{DS}}$. (a) and (b) show bias triangle pairs for $V_{\text{DS}} = +0.5~\text{mV}$ and $V_{\text{DS}} = -0.5~\text{mV}$ respectively. (c) and (d) show bias triangle pairs for $V_{\text{DS}} = +0.25~\text{mV}$ and $V_{\text{DS}} = -0.25~\text{mV}$ respectively.}
    \label{fig:Triangles_Zoom}
\end{figure}

To further investigate the formation and properties of the bias triangle pairs in our device, we performed transport measurements at $70~\text{mK}$, focusing on a specific bias triangle pair, see Fig.~\ref{fig:Triangles_Zoom}. The bias triangle pairs respond as expected with positive and negative $V_{\text{DS}}$, switching direction. See Fig.~\ref{fig:Triangles_Zoom}(a) and Fig.~\ref{fig:Triangles_Zoom}(b) for $V_{\text{DS}} = 0.5~\text{mV}$ and $V_{\text{DS}}=-0.5~\text{mV}$ respectively. The size of the triangles is also proportional to the magnitude of the applied $V_{\text{DS}}$. This is clear from comparing Fig.~\ref{fig:Triangles_Zoom}(a) with Fig.~\ref{fig:Triangles_Zoom}(c). 

\section*{Measurement Results: Single Electron Box\label{sec:measurement_results_seb}}

\begin{figure}[h]
    \centering
    \includegraphics[width=0.8\linewidth]{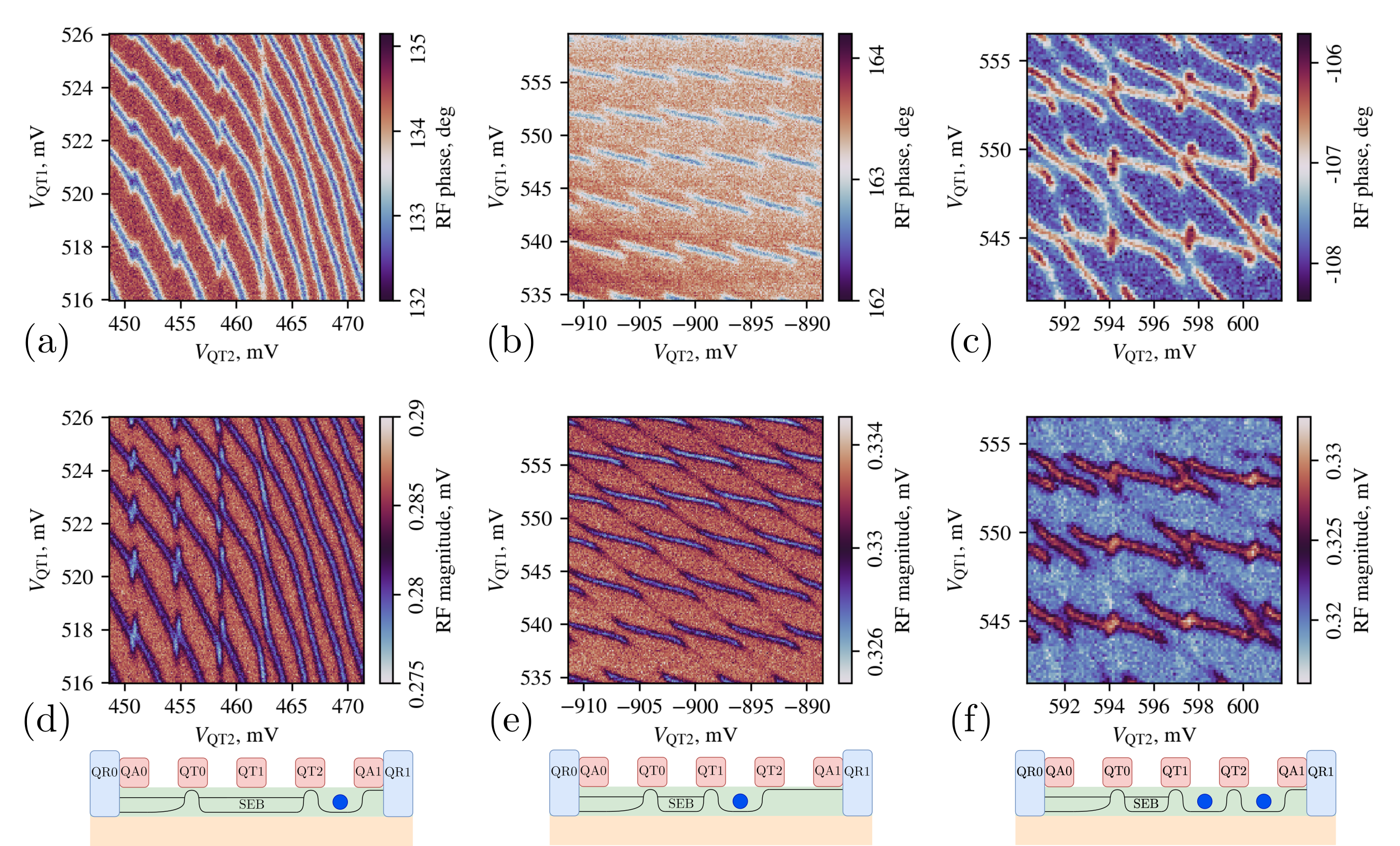}
    \caption{SEB measurement results for three different biasing conditions at $700~\text{mK}$. (a) and (d) show a large SEB forming between QT0 and QT2, with a quantum dot forming between QT2 and QA1. (b) and (e) show a SEB forming between QT0 and QT1, with a quantum dot forming between QT1 and QT2. (c) and (f) show a SEB forming between QT0 and QT1, with a double quantum dot forming between QT1, QT2, and QA1.}
    \label{fig:SEB_Fig_114_116}
\end{figure}

We finally present the formation of a SEB at the edge of our QDA structure, shown earlier to be compatible with integrated circuits~\cite{bashir2020single,petropoulos2024}. This experiment demonstrates its use in detecting charge transitions of a coupled quantum dot. We consider three cases; a large SEB is first formed between QT0 and QT2 sensing a single quantum dot, then a smaller SEB is formed between QT0 and QT1 sensing a single quantum dot, and finally, a SEB is formed between QT0 and QT1 sensing a double quantum dot. The phase and magnitude results of the reflectometry are included for all cases in Fig.~\ref{fig:SEB_Fig_114_116}. In biasing the SEB, we also ensure transport is blocked, by forming a barrier under QA1 in Fig.~\ref{fig:SEB_Fig_114_116}(a),(d) and under QT2 and QA1 in Fig.~\ref{fig:SEB_Fig_114_116}(b),(c),(e),(f).

In Fig.~\ref{fig:SEB_Fig_114_116}(a) we see the application of $V_{\text{QT2}}$ transforms the charge stability response from a double quantum dot to a single quantum dot with diagonal lines. Charge transitions are much more strongly coupled in all cases to variations in $V_{\text{QT1}}$ given the nature of SEB sensing. This is particularly clear in Fig.~\ref{fig:SEB_Fig_114_116}(b) where variations in $V_{\text{QT2}}$ show little to no response away from the triple points, whilst variations in $V_{\text{QT1}}$ show very distinct charge transition lines away from the triple points. At the triple points, distinct jumps are seen both in Fig.~\ref{fig:SEB_Fig_114_116}(a) and (b). As electrons are loaded into the quantum dot by the action of sweeping $V_{\text{QT2}}$, the response of the SEB shifts due to the electrostatic interaction between the SEB and the quantum dot, showing the characteristic shift in the charge stability transitions at the triple points~\cite{oakes_fast_2023}. The density of charge transitions in Fig.~\ref{fig:SEB_Fig_114_116}(a) is due to the larger SEB region, resulting in less spaced energy levels compared to the SEB forming in a smaller region in Fig.~\ref{fig:SEB_Fig_114_116}(b) and (c). 

The SEB response in the case of Fig.~\ref{fig:SEB_Fig_114_116}(c) and (d) is that of a triple quantum dot system, where the SEB is used to sense charge transitions in the coupled double quantum dot forming between QT1/QT2 and QT2/QA1. A number of transitions are more clearly seen in the phase response in Fig.~\ref{fig:SEB_Fig_114_116}(c) compared to the magnitude response in Fig.~\ref{fig:SEB_Fig_114_116} due to variations in the quantum capacitance of the double quantum dot that more strongly affect the phase response~\cite{vigneau_probing_2023}. A double quantum dot response is visible in Fig.~\ref{fig:SEB_Fig_114_116}(c) overlaid with another set of transitions due to the SEB. Further analysis is needed to understand each transition in detail in this case and to fit using an appropriate theoretical framework~\cite{gaudreau_stability_2006,lee_charge_2013}.

\section*{Conclusions}

We showed an experimental characterisation of a quantum dot array manufactured  in a commercial 22~nm Fully Depleted Silicon-On-Insulator process at GlobalFoundries. We presented a double-dot characterisation with a precise control of the barrier gate separating the two dots, as well as the formation of bias triangle pairs. In addition, we showed that the QDA can be configured with a single-electron box sensor at the edge of the array and be used to measure the charge stability of quantum dots forming in the same quantum dot array. 

\bibliography{references.bib}

\end{document}